\documentstyle[12pt,preprint]{aastex}
\begin{document}

\title{Spectral Variations of Of?p Oblique Magnetic Rotator Candidates in the Magellanic Clouds}

\author{Nolan R. Walborn}
\affil{Space Telescope Science Institute\altaffilmark{1}, 3700 San Martin Drive, Baltimore, MD 21218}
\authoremail{walborn@stsci.edu}

\author{Nidia I. Morrell}
\affil{Las Campanas Observatory, Carnegie Observatories, Casilla 601, La Serena, Chile}
\authoremail{nmorrell@lco.cl}

\author{Ya\"el Naz\'e\altaffilmark{2}}
\affil{GAPHE, D\'epartement AGO, Universit\'e de Li\`ege, All\'ee du 6 Ao\^ut 19c, Bat. B5C, B4000-Li\`ege, Belgium}
\authoremail{naze@astro.ulg.ac.be}

\author{Gregg A. Wade}
\affil{Department of Physics, Royal Military College of Canada, PO Box 17000 Station Forces, Kingston, ON, Canada K7K 7B4}
\authoremail{wade-g@rmc.ca}

\author{Stefano Bagnulo}
\affil{Armagh Observatory, College Hill, Armagh BT61 9DG, UK}
\authoremail{sba@arm.ac.uk}

\author{Rodolfo H. Barb\'a\altaffilmark{3}}
\affil{Departamento de F\'{\i}sica y Astronom\'{\i}a, Universidad de La Serena, Cisternas
1200 Norte, La Serena, Chile}
\authoremail{rbarba@dfuls.cl}

\author{Jes\'us Ma\'{\i}z Apell\'aniz\altaffilmark{3}}
\affil{Centro de Astrobiolog\'{\i}a, CSIC-INTA, Campus ESAC, Apartado Postal 78, E-28691 Villanueva de la Ca\~nada, Madrid, Spain}
\authoremail{jmaiz@cab.inta-csic.es}

\author{Ian D. Howarth\altaffilmark{4}}
\affil{Department of Physics and Astronomy, University College London,
Gower Street, London WC1E 6BT, UK}
\authoremail{idh@star.ucl.ac.uk}

\author{Christopher J. Evans\altaffilmark{4}}
\affil{UK Astronomy Technology Centre, Royal Observatory Edinburgh, 
Blackford Hill, Edinburgh EH9 3HJ, UK}
\authoremail{chris.evans@stfc.ac.uk}

\author{Alfredo Sota}
\affil{Instituto de Astrof\'{\i}sica de Andaluc\'{\i}a--CSIC, Glorieta de la
Astronom\'{\i}a s/n, 18008 Granada, Spain}
\authoremail{sota@iaa.es}

\altaffiltext{1}{Operated by the Association of Universities for Research in Astronomy, Inc., under NASA contract NAS5-26555.}

\altaffiltext{2}{FRS-FNRS Research Associate.}

\altaffiltext{3}{Visiting Astronomer, Las Campanas Observatory.}

\altaffiltext{4}{Visiting Astronomer, Anglo-Australian Observatory.}

\begin{abstract}
Optical spectroscopic monitoring has been conducted of two O stars in the Small and one in the Large Magellanic Cloud, the spectral characteristics of which place them in the Of?p category, which has been established in the Galaxy to consist of oblique magnetic rotators.  All of these Magellanic stars show systematic spectral variations typical of the Of?p class, further strengthening their magnetic candidacy to the point of virtual certainty.  The spectral variations are related to photometric variations derived from OGLE data by Naz\'e et al. (2015) in a parallel study, which yields rotational periods for two of them.  Now circular spectropolarimetry is required to measure their fields, and ultraviolet spectroscopy to further characterize their low-metallicity, magnetically confined winds, in support of hydrodynamical analyses.   
\end{abstract}


\section{Introduction}

The Of?p class in the Galaxy currently consists of five objects (Walborn et al. 2010).  The sometimes maligned question mark in their classification, introduced by Walborn (1972, 1973), was intended to denote doubt that they are normal Of supergiants, as the latter category was being interpreted at the time, because of several relatively subtle spectral peculiarities.  One of them is emission-line profiles narrower than those of the absorption lines, suggesting a circumstellar origin of the former in the peculiar stars, rather than photospheric as in the normal ones.  That morphological distinction was amply vindicated decades later, when 1--20~kG magnetic fields were discovered in all of the Of?p stars (Donati et al. 2006; Martins et al. 2010; Wade et al. 2011, 2012a, 2012b, 2015; Hubrig et al. 2008, 2011, 2013, 2015).  Because of obliquity between their magnetic and rotational axes, they display spectacular variations during their rotational cycles, which range from about a week to more than 50~years among the Galactic objects, likely determined by spindown as a function of field strength and age, due to angular momentum loss produced by coupling of the magnetic fields to the winds (``magnetic braking'').  That is, they are higher-mass analogues of the Ap and Bp stars, with even more spectacular behaviors because the fields channel the strong stellar winds from opposite hemispheres into a collision at the magnetic equator, forming a thin disk that presents varying aspects.  While on the one hand there is a phenomenological unity among the class because of the basic mechanism, on the other hand the different combinations of multiple parameters also lead to a remarkable diversity of detail.

This diversity among the small number of Galactic Of?p stars is not merely amusing; the ranges among their individual parameters provide ``separated-variable experiments" that illuminate the entire class.  Clearly further examples of the phenomenon are of considerable interest from that perspective.  The first Of?p star in the Magellanic Clouds was recognized in the SMC by Walborn et al. (2000--still before the discovery of their physical nature!); two more in each Cloud have been found since.  They offer the interesting opportunity to investigate magnetic fields and winds of massive stars at lower metallicities.  Here we present spectroscopic monitoring of the three objects known for long enough to accomplish that; two more have been discovered very recently by Massey et al. (2014) and are not further discussed in this paper, although they are included in the companion photometric paper by Naz\'e et al. (2015).  
     
\section{Observations}

Spectroscopic observations of AV\,220 and 2dFS\,936 in the SMC, and BI\,57 in the LMC (designations explained below), have been obtained with the Boller \& Chivens spectrograph attached to the 2.5m du Pont telescope at Las Campanas Observatory.  The Marconi No.~1 2048$\times$515 13.5$\mu$-pixel detector was in use with that instrument.  We used a 1200~lines~mm$^{-1}$ grating centered at 4700~\AA\ and had the slit width set to 150 microns, corresponding to 1\farcs26 on the sky and 1.75 pixels on the detector.  This instrumental configuration produces a resolution of $\sim$1.7~\AA\ as measured from the FWHM of the comparison lines.  The typical peak signal-to-noise ratio (S/N) per 2-pixel resolution element ranges from 150 to 200, with just a few cases significantly below or above.  All of these data were obtained under the technical auspices of the Galactic O-Star Spectroscopic Survey (GOSSS; Ma{\'\i}z Apell\'aniz et al. 2011), and they have been incorporated into the GOSSS database.  

Dome flats and bias frames were obtained during the afternoon prior to each observing night as well as twilight flats at sunset, and He-Ne-Ar comparison-lamp exposures were recorded before or after each target was observed.  Reductions were carried out using standard IRAF\footnote{IRAF is distributed by the National Optical Astronomy Observatory, which is operated by the Association of Universities for Research in Astronomy (AURA), Inc., under cooperative agreement with the  National Science Foundation.} longslit routines.  Flux calibration was derived from a set of standard stars observed during the same nights, and the data were then normalized for presentation here.

One observation of AV\,220, one of 2dFS\,936, and two of BI\,57 were obtained
with the Magellan Echellette (MagE; Marshall et al. 2008) at the Clay 6.5m telescope, with a $1\arcsec\times5\arcsec$ slit, yielding a resolution ranging from 0.7~\AA\ at 3200~\AA\ to 1.7~\AA\ at 10,000~\AA, as measured from the FWHM of the Th-Ar comparison lines.  The peak S/N per 3-pixel resolution element is 95 for AV~220, 215 for 2dFS~936, and 140 for BI~57.  Reductions were done with a combination of the IRAF {\it mtools} package (originally designed by Jack Baldwin to process Magellan Echelle--MIKE--data) and standard IRAF echelle routines, as described by Massey et al. (2012). 

We also present here one observation of 2dFS\,936 and one of BI\,57
obtained by IDH and CJE with the 2dF instrument attached to the Anglo-Australian Telescope (see Lewis et al. 2002; Evans et al. 2004).  One spectrogram of 2dFS\,936 was obtained by IDH with the ANU 2.3m telescope and Dual-Beam Spectrograph (DBS) at Siding Spring Observatory, described by Rodgers et al. (1988), using 1200~lines~mm$^{-1}$ gratings and E2V detectors windowed to 2048$\times$512 13.5$\mu$ pixels, which gave a nominal resolution of $\sim$1~\AA.  Also discussed here for AV\,220 are one observation from CASPEC at the 3.6m on ESO/La Silla (Walborn et al. 2000), and data obtained with the 8m VLT/FLAMES on ESO/Paranal by CJE (LR02/LR03 configurations as described by Evans et al. 2011).

\section{Results}

All of the Galactic Of?p stars display distinctive, periodic spectral variations to various degrees, which are reproducible across the rotational cycles for as long as they have been observed (Naz\'e et al. 2001, 2008, 2010; Walborn et al. 2004; Howarth et al. 2007).  The phases of their magnetic extrema correlate with those of the spectral variations, which may be described as maxima and minima in terms of the emission-line intensities. These variations include the apparent spectral types, with the earlier types due to emission filling at the He~I lines, so that the minima are likely more representative of the actual types.  In some objects, the C~III $\lambda$4650 emission lines, whose comparability to the N~III $\lambda$4640 was one of the original criteria for the class, disappear abruptly and completely, as if in an occultation.  In some objects, H$\alpha$ changes from a strong emission line to an absorption, and comparable profile changes occur in He~II $\lambda$4686.  Higher Balmer series members develop variable, narrow P~Cygni profiles within the broad stellar absorption lines.  Several Of?p stars display very small-amplitude light variations that also correlate with the projected magnetic field and emission lines, because scattering electrons believed to produce the photometric maxima are concentrated over the magnetic poles, corresponding to face-on aspects of the disks which maximize the emission-line strengths (Walborn et al. 2004; Howarth et al. 2007; Wade et al. 2011, 2012a, 2015).      

Here we report precisely these kinds of variations in the spectra of all three of the MC objects discussed, as well as correlations with their light variations, thus essentially establishing their Oblique Magnetic Rotator (OMR) nature in advance of any magnetic observations.  Once again, the ongoing power of morphology to discover and characterize phenomena prior to physical measurement and analysis is demonstrated.  

\subsection{AV 220 in the SMC}

\begin{deluxetable}{ccccccc}
\tablecolumns{7}
\tablewidth{0pc}
\tablecaption{Summary of spectroscopic observations of AV\,220 ordered by epoch}
\tablehead{
\colhead{Date}&\colhead{HJD-2450000.0}&\colhead{SpT}&
\colhead{EW $\lambda$4686}&\colhead{EW H$\beta$}&\colhead{Instrument}}
\startdata
20 Nov 1997&0772.5&O6.5&-1.53$\pm$0.05&-2.50$\pm$ 0.07&CASPEC\\
27 Sep/04 Oct 2004$^{a}$&3278.5&O6.5&-2.00$\pm$0.10&-2.75$\pm$0.10&FLAMES\\
03 Dec 2011 & 5898.6& O7 & -1.15$\pm$0.05&-1.20$\pm$0.10&B\&C\\
06 May 2012 &6053.9 &O7.5&-0.86$\pm$0.10&-0.96$\pm$0.10&B\&C\\
10 May 2012 & 6057.9& O7.5&-0.85$\pm$0.10&-0.75$\pm$0.10&B\&C\\
16 Mar 2013 & 6368.5&O6.5&-1.15$\pm$0.05&-1.70$\pm$0.10&B\&C\\
29 May 2013 & 6442.9& O6 &-1.35$\pm$0.05&-1.80$\pm$0.10&B\&C\\
16 Jun 2014 & 6824.9& O5.5&-2.02$\pm$0.05&-2.57$\pm$0.10&B\&C\\
18 Jun 2014 & 6826.9& O5.5&-2.00$\pm$0.05&-2.73$\pm$0.10&B\&C\\
28 Jul 2014 & 6866.8& O5.5&-2.10$\pm$0.05&-2.89$\pm$0.10&B\&C\\
09 Jan 2015 & 7031.5& O6 & -2.00$\pm$0.05&-2.90$\pm$0.10&MagE\\
\enddata
\tablenotetext{a}{This observation combines 4 individual spectrograms obtained 2004 Sept. 27 \& 28 (covering $\lambda\lambda$3980--4560) and Oct. 1 \& 4 
($\lambda\lambda$4510--5050); the Julian Dates have been averaged.}
\end{deluxetable}

\begin{figure}
\includegraphics[angle=270,width=1.2\textwidth]{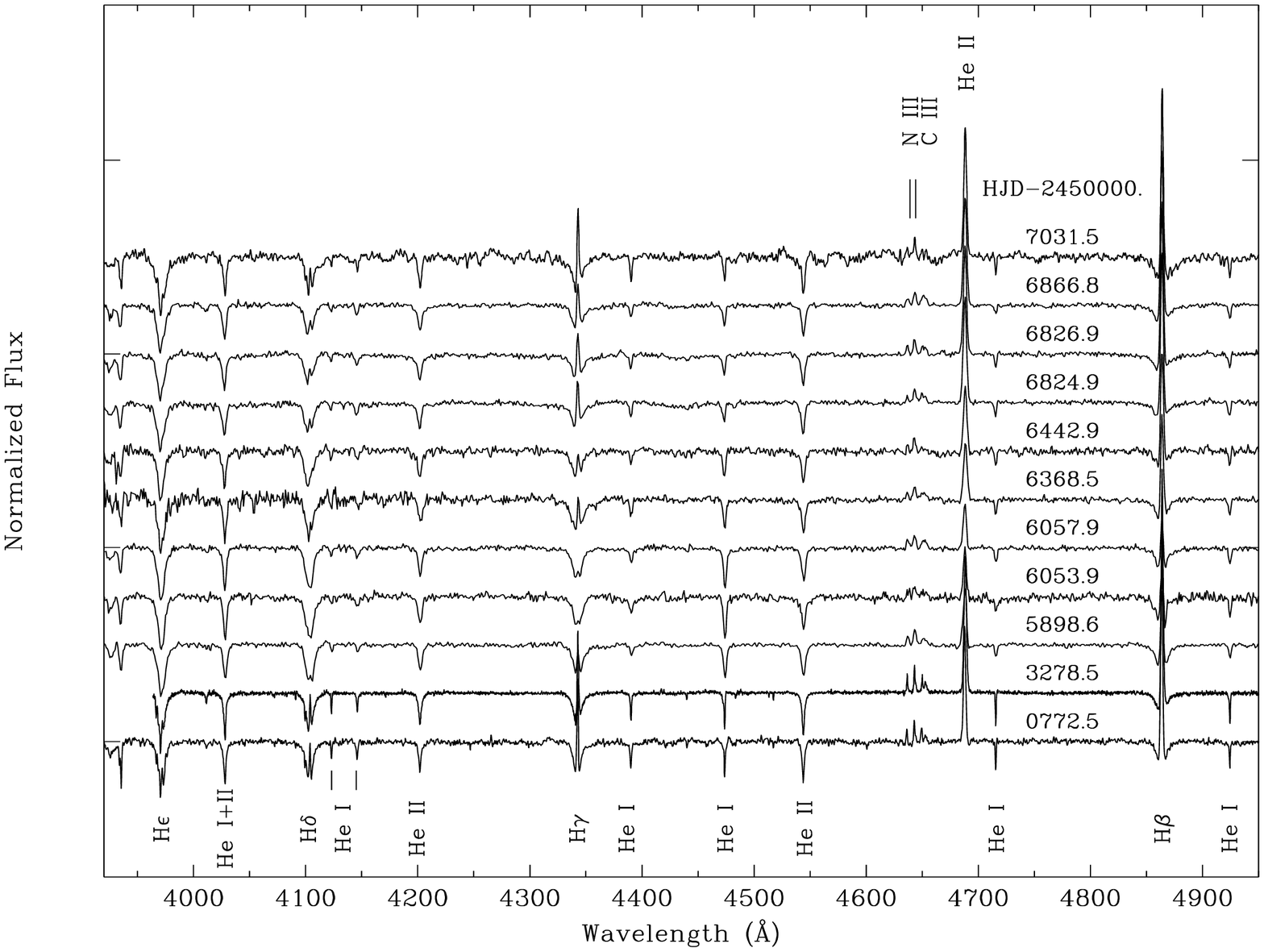}
\caption{Spectroscopic monitoring of AV 220 in the SMC.  Since there is no period for this object as yet, the data are plotted in chronological order.  Nevertheless, large variations typical of Of?p stars in spectral type (He~I/He~II absorption-line ratios) and emission-line intensities are seen (cf. Table~1).  Wavelengths of the identified lines may be found in the spectral atlas of Sota et al. (2011).}
\end{figure}

\begin{figure}
\epsscale{1.0}
\plotone{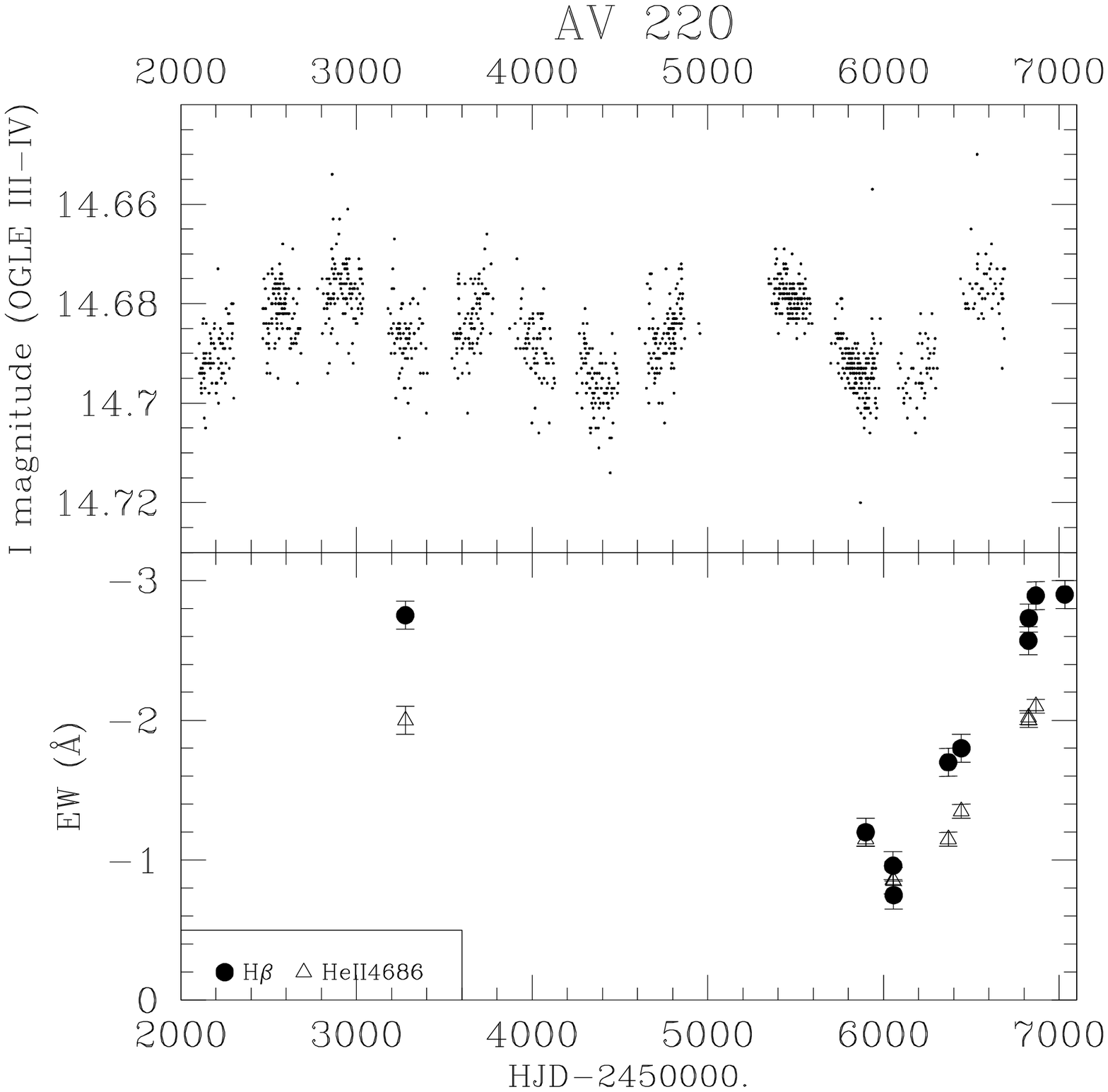}
\caption{Secular plot of the He~II $\lambda$4686 and H$\beta$ emission-line EWs in AV~220, compared with its lightcurve from Naz\'e et al. (2015), where details of the photometry can be found.  See text for discussion.}
\end{figure}

AV~220 is located just north of NGC~346, the largest H~II region in the SMC, but it is a field star sufficiently removed to be unaffected by nebular emission lines.  A chart is provided by Azzopardi et al. (1975, Chart~7).  AV~220 has $V$~=~14.50 and $M_V$~=~$-$4.9 (for $R$~=~3 and a distance modulus of 19.1), the latter far fainter than for a supergiant (Walborn 1973).  In fact, since the distances of most Galactic Of?p stars are uncertain, these MC objects provide valuable confirmation of their spectroscopic nonsupergiant interpretation.

Table~1 specifies and Figure~1 displays the 11 available spectrograms of AV~220.
The 1997 CASPEC data (Walborn et al. 2000, Fig.~6) comprise the first Of?p spectrum recognized in the Magellanic Clouds.  (Note also in Fig.~5 of that reference the high-quality {\it HST} UV observation made 24~May 1999, which displays very peculiar stellar-wind profiles and constitutes the only UV data for an MC Of?p star to date.)  

The apparent spectral type of AV~220 ranges from O5.5 to O7.5, i.e., the He~I~$\lambda$4471/He~II~$\lambda$4541 ratio strongly inverts about the unit value at O7.  The Balmer lines are highly variable, as best seen at H$\gamma$ in the figure, with central or redshifted emission ranging from very weak to very strong in correlation with the spectral type (which, again, varies due to emission in He~I $\lambda$4471; the weaker He~I lines are too strong for the earlier types, which is another characteristic of Of?p spectra no doubt related to the illusory maximum types).  The equivalent widths (EWs) of the He~II $\lambda$4686 and H$\beta$ emission lines (Table~1) vary correlatively with the spectral types.  On the other hand, any variations in the C~III/N~III emission ratio are at the limit of significance in these data; certainly the C~III never disappears.  The essential constancy of the spectrum from mid-June through July 2014 is indicative of a long period.

Unfortunately, despite an extensive database, clear variability, and exhaustive searches, the photometry has thus far not yielded a definitive period for this star (Naz\'e et al. 2015).  Therefore, it is not yet possible to relate the substantial spectroscopic variations to its (presumed) rotational phases.  Nevertheless, a secular plot of its lightcurve and emission-line EWs is presented in Figure~2, which shows an apparent correlation in the expected sense (i.e., weaker lines at a light minimum and viceversa) during the last $\sim$1000 days of overlapping coverage.  A timescale of roughly 1500 days could describe this segment.  However, the earlier photometry is not consistent with that, nor is the (multiple) spectroscopic observation at (average) HJD~2453278.5 (Table~1).  Evidently, another phenomenon in addition to rotation complicates the variations of AV~220.  

\subsection{2dFS 936 in the SMC}

\begin{deluxetable}{ccccccc}
\tablecolumns{7}
\tablewidth{0pc}
\tablecaption{Summary of spectroscopic observations of 2dFS\,936 ordered by phase}
\tablehead{
\colhead{Date}&\colhead{HJD-2450000.0}&\colhead{Phase$^{a}$}&\colhead{SpT}&
\colhead{EW $\lambda$4686}&\colhead{EW H$\beta$}&\colhead{Instrument}}
\startdata
15 Jun 2014 & 6823.9 & 0.07 & O7.5 & -2.43$\pm$0.10 &-0.65$\pm0.2$& B\&C\\
17 Jun 2014 & 6825.9 & 0.07 & O7.5 & -2.44$\pm$0.10 & -0.62$\pm0.2$&B\&C\\
28 Jul 2014 & 6866.8 & 0.10 & O7.5 & -2.75$\pm$0.10 & -0.78$\pm$0.2&B\&C\\
03 Sep 2014 & 6903.7 & 0.12 & O7   & -3.06$\pm$0.05 &-1.37$\pm$0.2& MagE\\
30 Sep 1999 & 1452.0 & 0.15 & O6.5 & -3.40$\pm$0.10 & ---&2dF\\
03 Dec 2011 & 5898.5 & 0.39 & O6.5 & -3.20$\pm$0.10 &-0.99$\pm$0.2& B\&C\\
05 May 2012 & 6052.9 & 0.50 & O6.5 & -3.31$\pm$0.10 &-1.42$\pm$0.2& B\&C\\
16 Mar 2013 & 6367.5 & 0.73 & O7.5 & -3.00$\pm$0.10 &-0.92$\pm$0.2& B\&C\\
31 Dec 2005 & 3736.0 & 0.80 & O7.5 & -2.60$\pm$0.05 & -0.98$\pm$0.2&DBS\\
\enddata
\tablenotetext{a} {Based upon P = 1370$\pm$30 days and $T_0$=2453993.3$\pm$1.7 (Table~2 of Naz\'e et al. 2015)} 
\end{deluxetable}

\begin{figure}
\includegraphics[angle=270,width=1.2\textwidth]{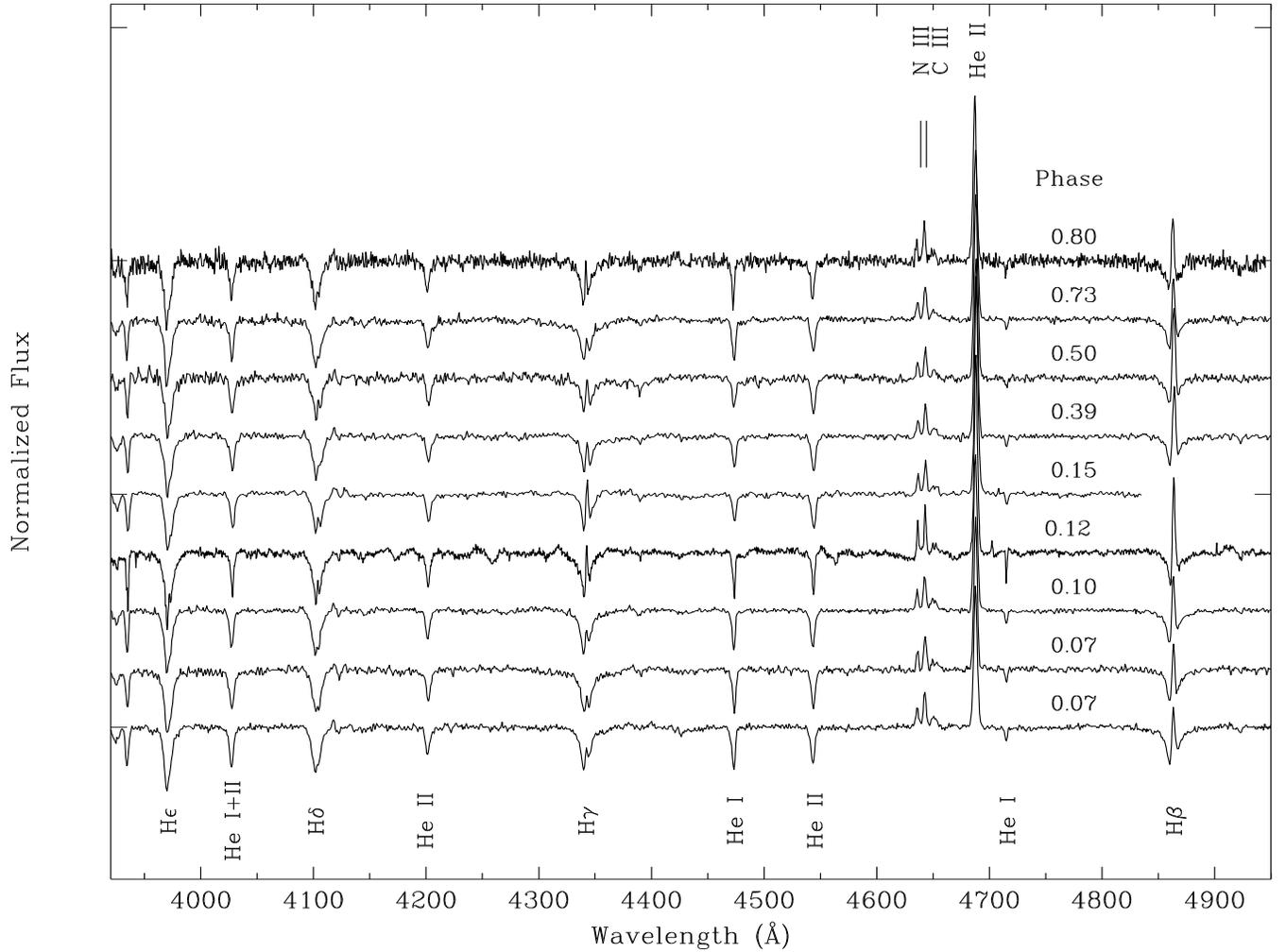}
\caption{Spectroscopic monitoring of 2dFS 936 in the SMC.  This star has a well determined photometric period of 1370$\pm$30 days so the spectroscopic data are ordered by phase (cf. Table~2 and discussion in the text).}
\end{figure}

\begin{figure}
\epsscale{1.0}
\plotone{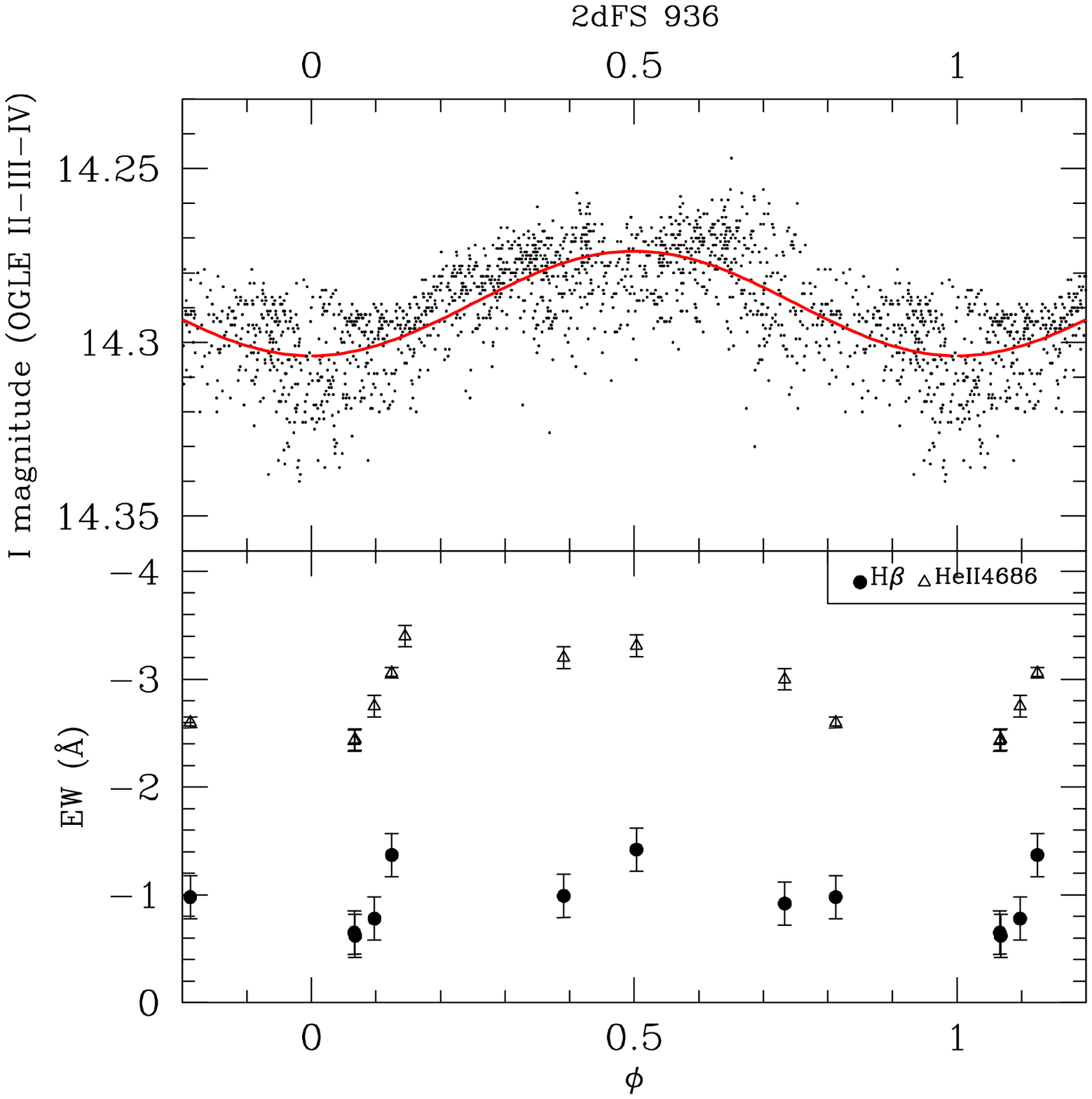}
\caption{Phase plot of the He~II $\lambda$4686 and H$\beta$ emission-line EWs in 2dFS~936, compared with its lightcurve from Naz\'e et al. (2015).}
\end{figure}

This object was first reported as Of?p by Massey \& Duffy (2001, Fig.~8), who provide a chart (their Fig.~6); it was subsequently observed and discussed by Evans et al. (2004, Fig.~17), whence the designation adopted here.  2dFS~936 has $V$~=~14.02 and $M_V$~=~$-$5.6, again fainter than a supergiant. 

Our observational and spectroscopic data for 2dFS~936, consisting of 9 spectrograms, are presented in Table 2 and Figure~3; they are ordered by phase as discussed below.  The apparent spectral type ranges from O6.5 to O7.5, with the He~I/He~II classification ratio again inverting through unity.  The Balmer emission at both H$\beta$ and H$\gamma$ varies significantly in intensity correlatively with the spectral types, but not in position, appearing always redshifted.  The C~III/N~III emission-line ratio is much smaller than in AV~220 and is essentially constant.  In view of the very long photometric period discussed next, the entire 2011--2014 spectroscopic sequence corresponds to a single rotational cycle.  

The EWs of the He~II $\lambda$4686 and H$\beta$ emission lines are listed in Table~2 and plotted in Figure~4 in comparison with the photometric lightcurve from Naz\'e et al. (2015), which corresponds to a well determined period of 1370$\pm$30 days.  Although sparse, the spectroscopic data show a tendency to track the lightcurve in the sense expected from Galactic counterparts, i.e., stronger emission lines near the light maximum.  The large ratio of the He~II to the Balmer line is noteworthy.     

\subsection{BI 57 in the LMC}

\begin{deluxetable}{cccccrrc}
\tablecolumns{8}
\tablewidth{0pc}
\tablecaption{Summary of spectroscopic observations of BI\,57 ordered by 400d phase}
\tablehead{
\colhead{Date}&\colhead{HJD-2450000.0}&\colhead{Phase$_1^{a}$}&
\colhead{Phase$_2^{b}$}&\colhead{SpT}&
\colhead{EW $\lambda$4686}&\colhead{EW H$\beta$}&\colhead{Instrument}}
\startdata
03 Dec 2004 & 3343.0 & 0.10 &0.51& O8 & -0.43$\pm$0.05 &-0.98$\pm$0.05& 2dF\\
09 Jan 2015 & 7031.7 & 0.32 &0.20& O7.5 &-0.55$\pm$0.05&-2.10$\pm$0.05& MagE\\
17 Feb 2013 & 6340.5 & 0.60 &0.32& O7.5& -0.50$\pm$0.10 &-1.74$\pm$0.10& B\&C \\
16 Mar 2013 & 6367.6 & 0.66 &0.35& O7.5-O8&-0.51$\pm$0.05&-1.56$\pm$0.05&B\&C \\
13 Feb 2012 & 5970.6 & 0.67 &0.85& O8 & -0.50$\pm$0.05 & -1.10$\pm$0.05&MagE \\
25 Apr 2014 & 6773.5 & 0.68 &0.87& O8 & -0.52$\pm$0.05&-0.97$\pm$0.05&B\&C \\
04 May 2012 & 6052.5 & 0.88 &0.95& O8  & -0.53$\pm$0.05 &-1.26$\pm$0.05& B\&C \\
28 Jul 2014 & 6866.9 & 0.91 &0.99& O8 & -0.60$\pm$0.05&-1.42$\pm$0.05&B\&C \\
\enddata
\tablenotetext{a} {Based upon P$_1$ = 400$\pm$3.5 days and $T_0$=2455302.2$\pm$1.2 (Table~2 of Naz\'e et al. 2015)}
\tablenotetext{b} {Based upon P$_2$ = 787$\pm$14 days and the same $T_0$ (Naz\'e et al. 2015)} 
\end{deluxetable}

\begin{figure}
\includegraphics[angle=270,width=1.2\textwidth]{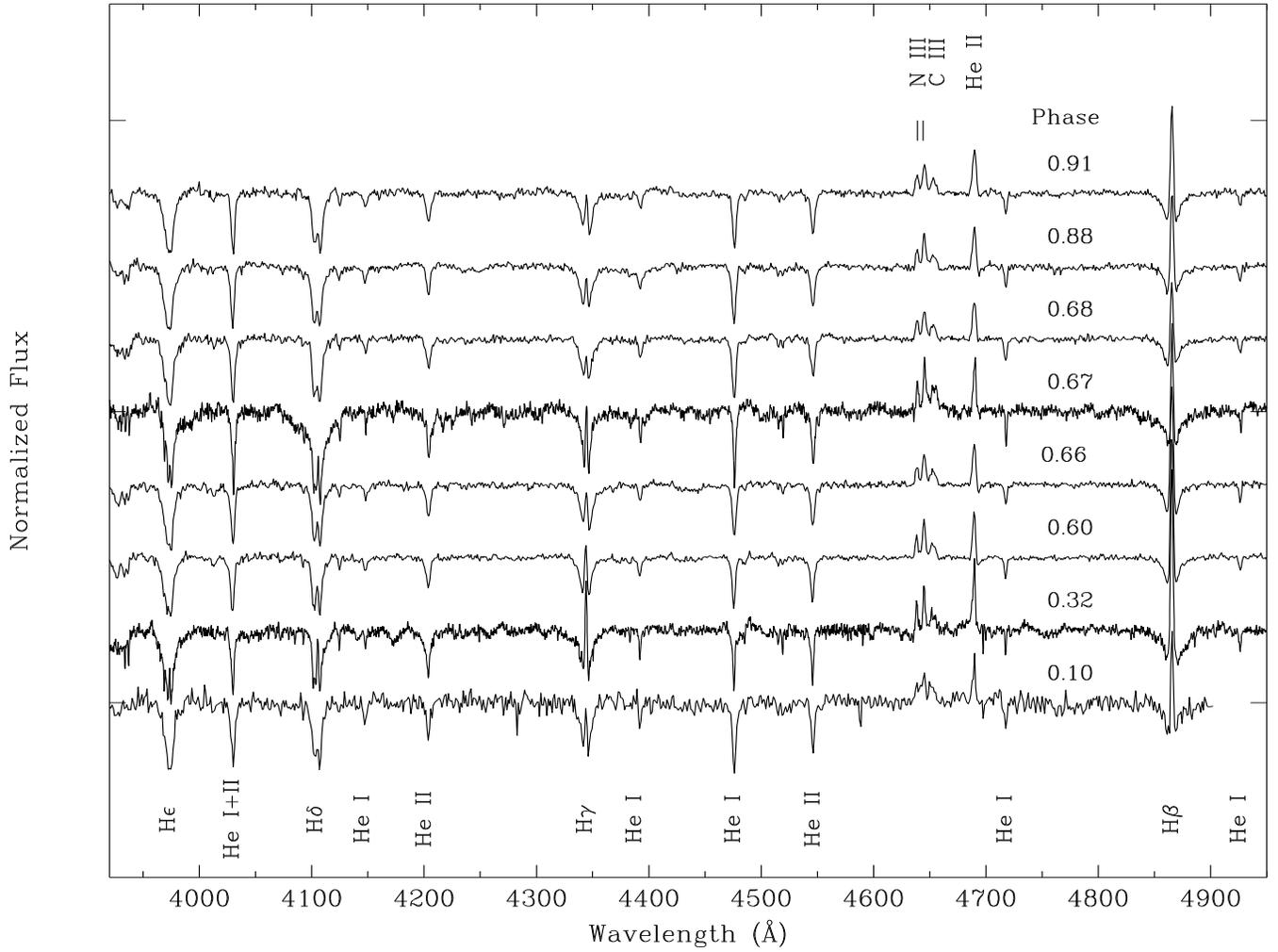}
\caption{Spectroscopic monitoring of BI 57 in the LMC.  The observations are ordered by phase corresponding to a period of 400$\pm$3.5 days (cf. Table~3 and discussion in the text).}
\end{figure}

\begin{figure}
\epsscale{1.0}
\plottwo{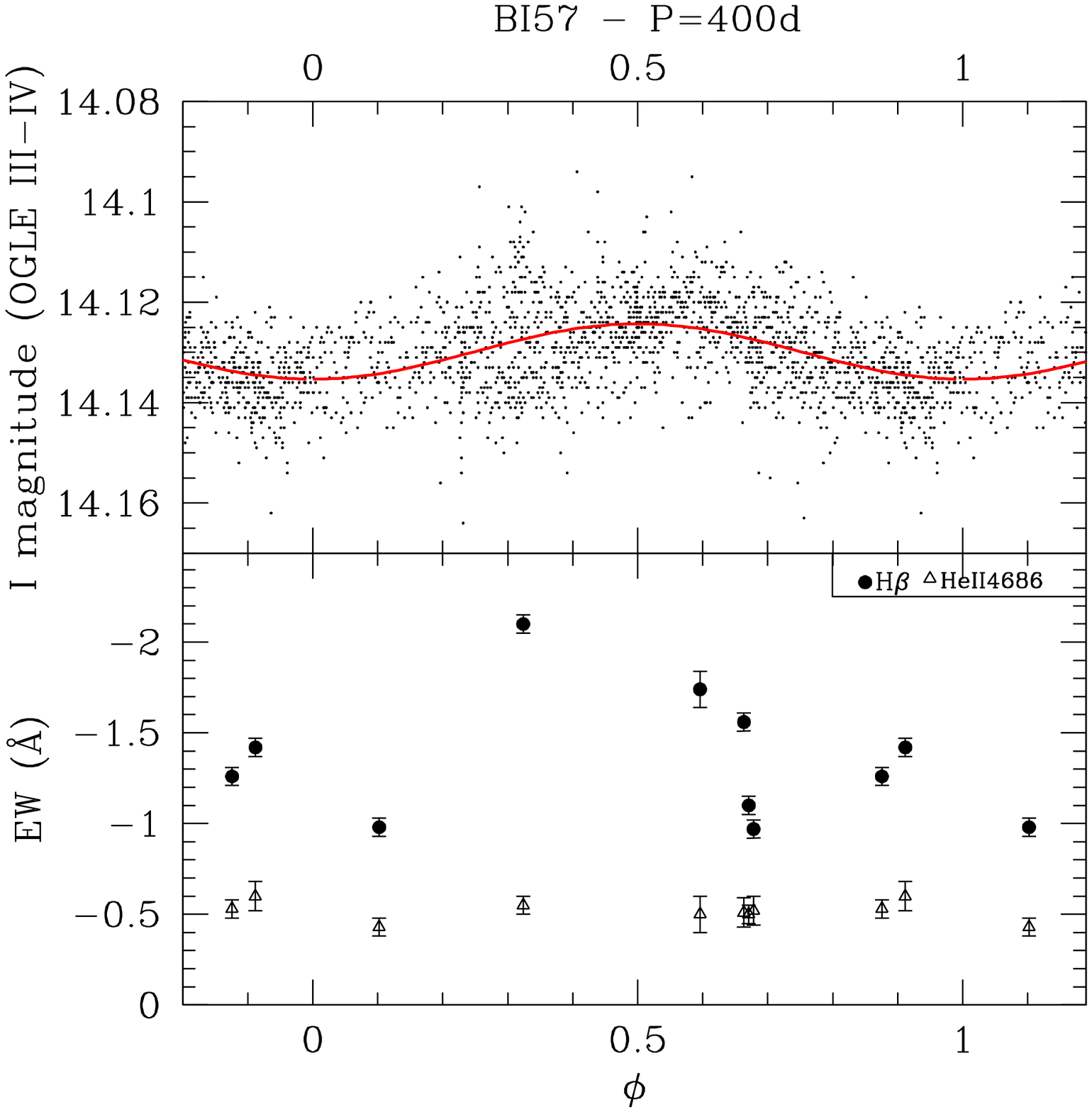}{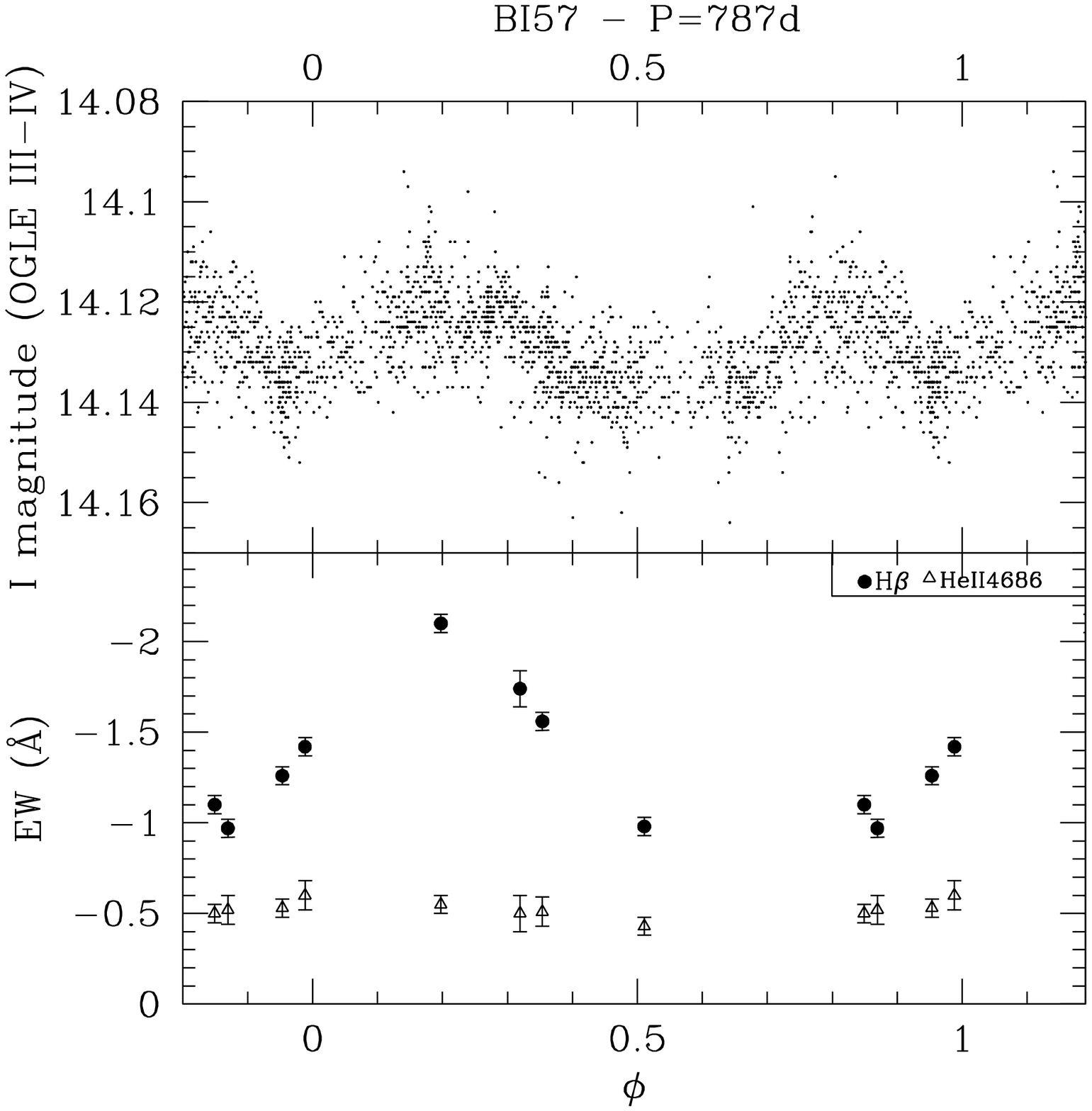}
\caption{Phase plots for 400d (left) and 787d (right) periods of the He~II $\lambda$4686 and H$\beta$ emission-line EWs in BI~57, compared with the corresponding lightcurves (Naz\'e et al. (2015).}
\end{figure}



The Of?p nature of this object was detected in an extensive, unpublished LMC survey by IDH.  It was subsequently identified with BI~57 (Brunet, Imbert et al. 1975), who provide a chart in their Plate~1 (second Hodge-Wright Atlas 28B cutout).  It has $V$~=~13.97 and $M_V$~=~$-$5.0 (for a DM of 18.6), similar to the other two stars above and again much fainter than a supergiant, as expected and now confirmed for Of?p stars. 

The data for 8 spectroscopic observations of this star are presented in Table~3 and Figure~5; the one at phase 0.10 of the 400d period is from the discovery survey.  The short spectral-type range is always later than O7, i.e., O7.5 to O8, which may seem small but is morphologically (and presumably physically) significant.  In this spectrum the variable Balmer emission is always either central or blueshifted, while there is a small but definite variation in the C~III/N~III emission ratio (compare $\lambda$4650 to $\lambda$4634).  On the other hand, the relatively weak He~II $\lambda$4686 emission is essentially constant whereas H$\beta$ is stronger and more variable (cf. EWs in Table~3), the opposite of the case for 2dFS~936.  In some spectra there is a weak absorption feature longward of the He~II $\lambda$4686 emission, creating the appearance of an inverse P~Cygni profile. 

Apart from the absolute magnitude and Of?p variations, there is no question of this spectrum arising from a normal (or even Ofc--Walborn et al. 2010) LMC supergiant.  At this late O type, there would be very strong Si~IV absorption lines flanking H$\delta$ that are not present, and the Balmer absorption itself is too strong for a supergiant.  Quantitative analyses of the high-resolution data for all three objects discussed here will be performed in due course.

Although not well distributed, the phase behavior of the He~II $\lambda$4686 (constant) and H$\beta$ emission EWs is compared with both the 400$\pm$3.5 and 787$\pm$14-day lightcurves found by Naz\'e et al. (2015) in Figure~6.  While there is some semblance of a correlation with H$\beta$ on 400d, it is not entirely satisfying.  The photometric searches found a second value of 787d with comparable probability from different methods and epoch ranges, i.e., essentially twice the first value.  Here two light maxima appear near phases 0.2 and 0.8, unequal particularly in their widths.  The correlation of the H$\beta$ EWs with the first maximum is remarkable, far better than on the 400d period; unfortunately, the second light maximum has inadequate spectroscopic coverage so any correlation is unclear.  If the 787d period is correct, then the 400d period is of course a harmonic of it; and it is likely that both magnetic poles are visible during the rotational cycle (as in the Galactic magnetic O stars HD~57682 and CPD~$-28^{\circ}$~2561; Grunhut et al. 2012, Wade et al. 2015, respectively), albeit unequally in this case.      

      
\section{Discussion and Outlook}

\subsection{Magnetic Observations}

From detailed spectroscopic monitoring of three MC Of?p objects, we have found distinctive variations analogous to those displayed by their Galactic counterparts, all of which have been established as OMRs.  Thus, with high confidence these MC objects also belong to that category and they are identified as the first extragalactic magnetic O stars in advance of any magnetic observations.  Nevertheless, it is of course essential to measure the strengths, orientations, and (aspect) variations of their fields directly, in order to support physical modeling; in particular, the inclination of the rotational axis and obliquity of the magnetic axis must be known to model the spectroscopic and light variations. The only current instrument capable of such measurements is FORS at the ESO VLT (Bagnulo et al. in preparation).  Since we have now derived rotational ephemerides for two of them (Naz\'e et al. 2015), we can predict the most favorable epochs for those observations, i.e., at the photometric and spectroscopic maxima.    

\subsection{Ultraviolet Observations}

The UV stellar-wind profiles and variations, especially in their blueshifted absorption troughs, provide vital information about the structure of the OMR magnetospheres.  As discussed above, the field channels the winds from opposite hemispheres into a collision at the magnetic equator, where high-energy phenomena are produced.  As the magnetosphere rotates rigidly and obliquely, the variable wind profiles diagnose the structure and flows in different directions. Indeed, three Galactic Of?p stars observed at multiple phases with {\it HST} display very peculiar and variable UV wind profiles, which can be understood in terms of magnetohydrodynamic models (Marcolino et al. 2012, 2013; Naz\'e et al. submitted).  A further Galactic object is currently under observation (Petit et al. in preparation).  As also mentioned above, a single {\it HST} observation of AV~220 (Walborn et al. 2000) likewise exhibits peculiar wind profiles, but it is the only extant UV observation of an MC Of?p star.  Now it is essential to obtain UV observations at the magnetic and optical extrema of the two objects with periods discussed here, and to pursue the period of the third for that purpose, to enable full characterization of their magnetospheric structures.

\acknowledgments

We thank Phil Massey for MagE observations obtained during joint observing runs.
NIM also thanks the LCO technical staff for excellent support during her runs. 
An anonymous referee provided some useful comments that improved the presentation.  YN acknowledges support from the Fonds National de la Recherche Scientifique (Belgium), the PRODEX XMM contracts, and the ARC (Concerted Research Action) grant financed by the Federation Wallonia-Brussels.  GAW acknowledges Discovery Grant support from the Natural Science and Engineering Research Council (NSERC) of Canada.  RHB acknowledges support from FONDECYT (Chile) via Regular Grant No.~1140076.  JMA and AS acknowledge support from (1) the Spanish Government Ministerio de Econom{\'\i}a y Competitividad (MINECO) through grants AYA2010-15\,081, AYA2010-17\,631, and AYA2013-40\,611-P; and (2) the Consejer{\'\i}a de Educaci{\'o}n of the Junta de Andaluc{\'\i}a through grant P08-TIC-4075.  Page charges were covered by NASA through grant HST-GO-13629.003-A (scientific PI YN) from STScI.

\end{document}